\documentclass[notitlepage,nofootinbib,onecolumn,11pt,a4paper,showpacs,preprintnumbers,amsmath,amssymb]{revtex4-1}
\date{March 2010}
\usepackage{epsfig}
\usepackage{latexsym}
\usepackage{amssymb}
\usepackage{booktabs}
\usepackage{graphicx}
\newcommand{\be}{\begin{equation}}
\newcommand{\ee}{\end{equation}}
\newcommand{\ba}{\begin{eqnarray}}
\newcommand{\ea}{\end{eqnarray}}
\newcommand{\bi}{\begin{itemize}}
\newcommand{\ei}{\end{itemize}}
\newcommand{\ud}{\,\mathrm{d}}
\newcommand{\tr}{{\rm Tr\,}}
\newcommand{\re}{\mathop{\rm Re}}

\newcommand{\<}{\langle}
\renewcommand{\>}{\rangle}
\newcommand{\eq}{Eq.~}

\newcommand{\fig}{Fig.~}

\newcommand{\la}{\label}

\newcommand{\txts}{\textstyle}

\newcommand{\im}{\mathop{\rm Im}}

\newcommand{\Nt}{N_{\tau}}
\newcommand{\Ns}{N_{\sigma}}

\newcommand{\bx}{\boldsymbol{x}}
\newcommand{\bp}{\boldsymbol{p}}
\newcommand{\bk}{\boldsymbol{k}}

\begin{document}
\preprint{MKPH-T-10-39}
\title{The errant life of a heavy quark in the quark-gluon plasma}

\author{Harvey~B.~Meyer}
\affiliation{Johannes Gutenberg Universit\"at Mainz, 
    Institut f\"ur Kernphysik, 55099 Mainz, Germany}

\date{\today}

\begin{abstract}
In the high-temperature phase of QCD, the heavy quark momentum
diffusion constant determines, via a fluctuation-dissipation relation,
how fast a heavy quark kinetically equilibrates.  This transport coefficient can
be extracted from thermal correlators via a Kubo formula.  We present
a lattice calculation of the relevant Euclidean correlators in the
gluon plasma, based on a recent formulation of the problem in
heavy-quark effective field theory (HQET).  We find a $\approx20\%$
enhancement of the Euclidean correlator at maximal time separation as
the temperature is lowered from $6T_c$ to $2T_c$, pointing to stronger
interactions at lower temperatures. At the same time, the correlator becomes flatter from
$6T_c$ down to $2T_c$, indicating a relative shift of the spectral
weight to lower frequencies. A recent next-to-leading order
perturbative calculation of the correlator agrees with the
time dependence of the lattice data at the few-percent level.  We
estimate how much additional contribution from the
$\omega\lesssim T$ region of the perturbative spectral function would
be required to bring it in agreement with the lattice data at
$3.1T_c$.
\end{abstract}

\pacs{ 12.38.Gc, 12.38.Mh, 25.75.-q}
\maketitle

\section{Introduction}

Imagine a heavy quark moving through the plasma of light quarks and
gluons with more than its fair share of thermal kinetic energy
$\frac{3}{2}k_BT$.  Interactions with the medium will gradually cause
it to slow down, at a rate determined by the drag coefficient
$\eta$. It also experiences stochastic interactions which balance the
drag force in such a way that the heavy quark's kinetic energy tends
to $\frac{3}{2}k_BT$.  In this article we study from first principles
the strength of these stochastic interactions, which is characterized
by a parameter called the momentum diffusion coefficient and denoted
by $\kappa$.  This is a quantity of phenomenological interest.
Measurements at the Relativistic Heavy Ion Collider (RHIC) have shown
that heavy quarks display substantial elliptic
flow~\cite{Abelev:2006db,Adare:2006nq}, implying stronger medium
interactions than extrapolated weak-coupling calculations would
suggest.  It has been estimated~\cite{CasalderreySolana:2006rq} that a
diffusion coefficient $D\lesssim 1/T$, or equivalently,
$\kappa/T^3\gtrsim 2$, is required in order to accomodate the RHIC
data. Theoretically, we will be concerned with collisional 
energy loss, which is expected to dominate for small $P/M$; however
one should keep in mind that 
in heavy-ion collisions, the quarks are not produced 
at rest, hence radiative energy loss can play a role too -- see
\cite{Renk:2010su} for a recent review.
A separate motivation is that the dynamics of the heavy-quark
probe is somewhat simpler to study than the diffusion of transverse
momentum carried by the constituents of the plasma, which is
characterized by the shear viscosity.  There are reasons to expect
that once the heavy quark diffusion constant has been determined, it
can be used to calibrate other transport coefficients such as the
shear viscosity, because ratios of transport coefficients are more
stable predictions of the weak-coupling expansion~\cite{Moore:2004tg}.

We briefly review the known analytic results for the momentum diffusion coefficient.
To strict leading order, $\kappa$ is given by~\cite{Moore:2004tg}
($C_F\equiv \frac{N_c^2-1}{2N_c}$)
\be \kappa = \frac{g^2 C_F T}{6\pi} m_D^2
\left(\log \frac{2T}{m_D}+\frac{1}{2}-\gamma_E +
\frac{\zeta'(2)}{\zeta(2)} +\frac{N_{\rm f}\log 2}{2N_c+N_{\rm
    f}}\right). 
 \ee 
Unfortunately, for realistic values of the Debye screening mass $m_D$,
this expression is negative. Carrying out the integrals without
expanding in $m_D$ ($m_D^2=g^2T^2(N_c/3+N_{\rm f}/6)$ in leading
order), positive values are obtained, for instance at $g^2\approx3$,
$\kappa/T^3\approx 0.4$ for $N_{\rm f}=3$ and 0.23 for $N_{\rm
  f}=0$~\cite{Moore:2004tg}.  Remarkably, a next-to-leading order
computation has been carried out~\cite{CaronHuot:2007gq}, with the
result $\kappa/T^3\approx 2.7$ for $g^2\approx3.0$ for $N_{\rm f}=3$.
It is unclear at this point how useful the expansion is.

The ${\cal N}=4$ SYM theory provides an interesting testing ground for
analytic methods. The result obtained by holographic methods in the
strong coupling limit of ${\cal N}=4$ SYM theory is $\kappa/T^3 = \pi
\sqrt{\lambda}$~\cite{CasalderreySolana:2006rq,Herzog:2006gh,Gubser:2006bz},
where $\lambda\equiv g^2N_c$ is the 't Hooft coupling.  On the other
hand, the NLO weak-coupling result has also been worked
out~\cite{CaronHuot:2008uh},
\be
\kappa^{\rm SYM} = \frac{\lambda^2 T^3}{6\pi}
\left(\log\frac{1}{\sqrt{\lambda}} + 0.4304 + 0.8010\sqrt{\lambda}\right).
\la{eq:NLO_SYM}
\ee
Although the apparent convergence is again poor, it is encouraging
that the strong coupling result crosses the weak-coupling result at an
intermediate coupling of $\lambda\approx10.2$.  This suggests that the
values obtained from NLO expressions such as (\ref{eq:NLO_SYM}) are in
the right ball-park\footnote{$\lambda\approx10.2$ is a moderately
  large coupling.  For a 't Hooft coupling of 10--12, the spatial
  scalar correlator computed by holographic methods in the SYM theory
  and on the lattice in the SU($N_c$) gauge theory exhibits a
  non-trivial agreement up to about $2T_c$~\cite{Iqbal:2009xz}.  In
  this range, the O($1/\lambda^{3/2}$) correction to the
  $\lambda=\infty$ shear viscosity to entropy density ratio amounts to
  about $20\%$.}.

In the next section, we review the steps that lead to a Kubo formula
for the momentum diffusion coefficient $\kappa$ in the static limit 
of the probe-quark. In section III we present our numerical results 
for the Euclidean correlators obtained by Monte-Carlo simulations,
and conclude in section IV.

\section{The spectral function of the heavy-quark current}

This section is mainly a review of recent literature on the subject of
heavy-quark diffusion, with some comments pertinent to calculations
performed in Euclidean space.  The basic definitions of thermal
correlators and the relations among them are gathered in the Appendix,
to which we refer the reader for unexplained notation. The main
equations of the linear response framework are also reviewed there.

To study the diffusion of a conserved charge such as the heavy quark
number $N=\int\ud\bx\, n(t,\bx)$, the Hamiltonian can be perturbed by
\be
H_{\mu} = H - {\int} \ud\bx\, \mu(t,\bx) n(t,\bx),\qquad\quad
\mu(t,\bx) = \mu(\bx)\,e^{\epsilon t} \theta(-t).
\ee
In the hydrodynamic treatment of the problem, Fick's law $\boldsymbol{j} = -D\nabla n$ 
and the conservation equation $\partial_t n + \nabla\cdot n = 0$ lead to the 
diffusion equation, whose solution in Fourier space takes the form 
\be
\tilde n(\omega,\bk) = \frac{\chi_s(\bk) \mu(\bk)}{-i\omega +D\bk^2},\qquad
\tilde n(\omega,\bk)\equiv \int_0^\infty e^{i\omega t} \int \ud\bx \,e^{-i\bk\cdot\bx}n(t,\bx).
\ee
Here $\chi_s(\bk) = \beta \int \ud\bx \,e^{-i\bk\bx}\< n(t,\bx) n(0)\>$ 
and $\chi_s\equiv \chi_s(\boldsymbol{0})$ is the particle number susceptibility.
Via (\ref{eq:retresp}), this determines the retarded correlator $G^{nn}_R(\omega)$
for small $\omega$ and $\bk$,
\be
G_R^{nn}(\omega,\bk) = \frac{(D\bk^2)^2 + i\omega\, D\bk^2}
                      {\omega^2+(D\bk^2)^2} \,\chi_s(\bk).
\ee
Before proceeding further, it is also instructive to write down the correlator in the real-time domain, 
\be
G_R^{nn}(t,\bk) \stackrel{t\to\infty}{\sim} \chi_s(\bk) D\bk^2 \, \exp\big(-D\bk^2 t\big).
\ee
Using \eq(\ref{eq:GEG}), the contribution of this exponential tail to
the Euclidean correlator (say) at $t=\beta/2$ is $\chi_s(\bk)$,
independently of the diffusion coefficient $D$.  This is a
manifestation of the difficulty to extract transport information from
Euclidean correlators.  

The longitudinal part of the current
correlator (i.e.$\<j_zj_z\>$ if $\bk=(0,0,k)$) is related to the
density correlator by the current conservation equation, $\rho_L(\omega,\bk) =
\frac{\omega^2}{\bk^2} \rho^{nn}(\omega,\bk)=\frac{1}{\pi}\frac{\omega^2}{\bk^2} \im G_R^{nn}(\omega,\bk)$. 
Thus the current spectral function reads
\be
\frac{\rho_L(\omega,\bk)}{\omega} =
\frac{\chi_s(\bk)}{\pi} \frac{ D\omega^2}{\omega^2+(D\bk^2)^2},
\la{eq:rhoLhydro}
\ee
implying in particular the Kubo formula
\be
D\chi_s = \pi\lim_{\omega\to0}\lim_{\bk\to0} \frac{\rho_L(\omega,\bk)}{\omega}.
\la{eq:kubo}
\ee

The diffusion of a heavy quark (one has in mind the charm or preferably, from 
a theoretical point of view, the bottom quark)
in the quark-gluon plasma is characterized by a time scale $M/T^2$
which is long compared to the thermal time scale of $1/T$.  For this
reason it is expected that a classical Langevin equation should
appropriately describe the thermalization of heavy quarks~\cite{Moore:2004tg}.  
See~\cite{Son:2009vu} and References therein for a derivation.
The heavy-quark's classical equations of motion are 
\ba 
\frac{d\bx}{dt} = \frac{\bp}{M},
&\quad & \frac{d\bp}{dt} = \boldsymbol{\xi}(t) - \eta \bp(t), 
\\ \<\xi^{i}(t) \xi^{j}(t')\> &=&
\kappa \delta^{ij} \delta(t-t').  
\la{eq:xixi}
\ea 
For a given $\xi(t)$, the equation is easily solved to give
\be
\bp(t) = e^{-\eta t} \big[\bp(0) +
 {\txts\int_0^t} \ud s \boldsymbol{\xi}(s) e^{\eta s}\big]\,,
\ee
implying
\be
\lim_{t\to\infty} \< p_i(t) p_j(t)\> = \frac{\kappa}{2\eta} \delta_{ij}.
\ee
The equipartition of energy requires $\frac{\bp^2}{2M}$ to be $\frac{3}{2}T$ in
equilibrium; the drag and fluctuation coefficients are thus related by the
fluctuation-dissipation relation 
\be
\eta = \frac{\kappa}{2MT}.  
\la{eq:Einstein}
\ee 
The mean square distance covered by the particle is also easily worked out.
For a thermal initial distribution of momenta, $\<p_i(0) p_j(0)\>=MT\delta_{ij}$, 
it reads
\be
{\txts\frac{1}{3}}\<\bx^2(t)\> = 
2D\big[ t  - {\txts\frac{1}{\eta}}(1-e^{-\eta t})\big].
\la{eq:x2lange2}
\ee
This equation describes both the early-time directed motion,
$ {\txts\frac{1}{3}}\<\bx^2(t)\> = \frac{1}{3} v^2 t^2$, ${\txts\frac{1}{3}}v^2=\frac{T}{M}$, 
and the late-time diffusive motion, ${\txts\frac{1}{3}}\<\bx^2(t)\> = 2Dt$~\cite{Petreczky:2005nh}.
Let now $P(t,\bx)$ be the probability that a heavy quark starts at the
origin at $t = 0$ and moves a distance $\bx$ over a time $t$.
If the distribution of heavy quarks at time zero is $n(0,\bx)$, 
at time $t$ it will be given by the convolution
\be
n(t,\bx) = \int \ud\bx' \, P(t,\bx-\bx')\, n(0,\bx'),
\ee
or equivalently 
\be
n(t,\bk) = P(t,\bk) \,n(0,\bk).
\ee
If one assumes the noise to be Gaussian distributed, 
then the probability distribution $P(t,\bx)$ is
Gaussian~\cite{Petreczky:2005nh}, with a width given
by \eq(\ref{eq:x2lange2}), and therefore so is $P(t,\bk)$. 
Applying the general rule (\ref{eq:retresp}), one then obtains
the retarded correlator,
\be
G^{nn}_R(\omega,\bk) = \chi_s(\bk) 
\left[ 
1+i\omega \int_0^\infty \ud t \, e^{i\omega t}\, P(t,\bk)
\right]
\la{eq:G_HQ}
\ee
The resulting spectral functions were obtained numerically in~\cite{Petreczky:2005nh}.
In particular, the $\bk=0$ spectral function takes the form
\be
\frac{\rho_L(\omega,\boldsymbol{0})}{\omega}= 
\frac{\chi_s}{\pi}\frac{T}{M}\,\frac{\eta}{\omega^2+\eta^2}.
\la{eq:rho_L}
\ee

The spectral structure that is obtained from the Langevin equation is
expected to arise for a sufficiently heavy diffusing particle.
Conversely, the presence of a transport peak allows one
to \emph{define} the quantities appearing in the Langevin equation
directly from the spectral function~\cite{CaronHuot:2009uh}. The
effective mean-square velocity is then given by
\be
{\txts\frac{1}{3}}\<\boldsymbol{v}^2\> \equiv \frac{1}{\chi_s}\int_{-\Lambda}^{\Lambda} 
 \frac{\ud\omega}{\omega} \rho_L(\omega),
\ee
where $\Lambda$ is a cutoff that separates the scale $\eta$ from 
the correlation time of the medium (which is typically of order $T$, or $gT$ at weak coupling).
The `kinetic mass' $M_{\rm kin}$ is further defined so as to satisfy the 
equipartition theorem, $M_{\rm kin}\<\boldsymbol{v}^2\>=3T$.
Finally, the momentum diffusion coefficient $\kappa(M)$ can be defined as 
\be
\kappa(M)= \frac{2\pi M_{\rm kin}^2}{\chi_s} 
\;\omega\rho_L(\omega)\Big|_{\eta\ll |\omega| \ll \Lambda}.
\la{eq:kappa_M}
\ee
A weak-coupling calculation shows that while $\kappa(M)$ and $D$ are
only weakly dependent on $M$, the drag coefficient $\eta\sim
T^2/M\times $ a power of the coupling constant is parametrically small
compared to the medium time-scale.  By reexpressing the right-hand side of
(\ref{eq:kappa_M}) in terms of the correlator of two heavy-quark
currents, the authors of~\cite{CasalderreySolana:2006rq,CaronHuot:2009uh} were able to
formulate the task of computing $\kappa$ in the static limit of
Heavy-Quark Effective Theory (HQET). In that limit, the leading
contribution is the `force-force' correlator, where the force is given
by the Lorentz expression $g\boldsymbol{E}$.  The relevant
Euclidean correlator reads, after evaluating the fermion line
contractions,
\be
G^{\rm HQET}_E(t) = 
\frac{\Big\<\re\tr \big( U(\beta,t)gE_k(t,\boldsymbol{0})
                   U(t,0)gE_k(0,\boldsymbol{0})\big)\Big\>}
{ -3\; \< \re\tr U(\beta,0)\>},
\la{eq:G_HQET}
\ee
where the color parallel transporters $U(t_2,t_1)$ 
in the fundamental representation are propagators of static quarks.
In particular the Polyakov loop appears in the denominator of (\ref{eq:G_HQET}).
The momentum diffusion coefficient is given by the low-frequency limit
of the corresponding spectral function via \eq(\ref{eq:ClatRho}),
\be
\kappa = \lim_{\omega\to0} \frac{2\pi T}{\omega} \rho^{\rm HQET}(\omega).
\la{eq:kappa_HQET}
\ee
The obvious advantage of this formulation is that the large scale $M$
has disappeared from the problem. The spectral function $\rho^{HQET}$
has been studied in detail at next-to-leading order (NLO) in
perturbation theory~\cite{Burnier:2010rp}.  Remarkably, even in the
weak-coupling limit, the function is smooth as small frequencies. This
is in contrast with the narrow transport peaks that are found at
weak-coupling in e.g.~the shear channel. This property represents a clear
advantage for numerical studies of the spectral function: 
the form of the kernel in \eq(\ref{eq:ClatRho}) makes the Euclidean
correlator very insensitive to the functional form of the spectral
function at $\omega\lesssim T$.
On the downside, while the spectral function of the current-current
correlator grows as $\frac{N_c}{12\pi^2}\omega^2$ at large frequencies, the spectral function
of the $\boldsymbol{E}$-field correlator grows even faster, 
like $\frac{g^2C_F}{6\pi^2}\omega^3$~\cite{CaronHuot:2009uh}.
This implies that the low-frequency part makes a comparatively small contribution to 
the Euclidean correlator studied in the next section.

\section{Force-force correlators from lattice gauge theory}

In this section we describe a first calculation of the HQET
`force-force' correlator (\ref{eq:G_HQET}) by means of Monte-Carlo
simulations in Euclidean space. The calculation is performed in the
deconfined phase of SU(3) gauge theory.  We employ the isotropic
Wilson action~\cite{Wilson:1974sk},
\be
 S_{\rm g} =  \frac{1}{g_0^2} \sum_{x,\mu\neq\nu} \tr\{1-P_{\mu\nu}(x)\}\,,
\la{eq:Sg}
\ee
where the `plaquette' $P_{\mu\nu}$ is the product of four link
variables $U_\mu(x)\!\in\,$SU(3) around an elementary cell in the
$(\mu,\nu)$ plane.  The size of the lattice is $\Nt\times \Ns^3$, with 
periodic boundary conditions in all directions.
As a local update algorithm, we use the standard
combination of heatbath and
over-relaxation~\cite{Creutz:1980zw,Cabibbo:1982zn,Kennedy:1985nu,Fabricius:1984wp}
sweeps in a ratio increasing from 5 to 7 as the lattice spacing is
decreased.  No multi-level
algorithm~\cite{Luscher:2001up,Meyer:2002cd} was used here, although
we expect that for sufficiently small lattice spacing, such an
algorithm will be beneficial, since the fluctuations of the $\boldsymbol{E}$ fields
will be UV-dominated. To set the scale we use the parametrization
of the Sommer scale $r_0\approx0.5$fm given in~\cite{Durr:2006ky}
based on the data~\cite{Necco:2001xg}, and convert between $r_0$ and $T_c$ 
by using $r_0T_c=0.746(7)$ (\cite{Meyer:2008tr} and Refs.~therein).

\begin{figure}
\centerline{\includegraphics[width=9.0 cm,angle=-90]{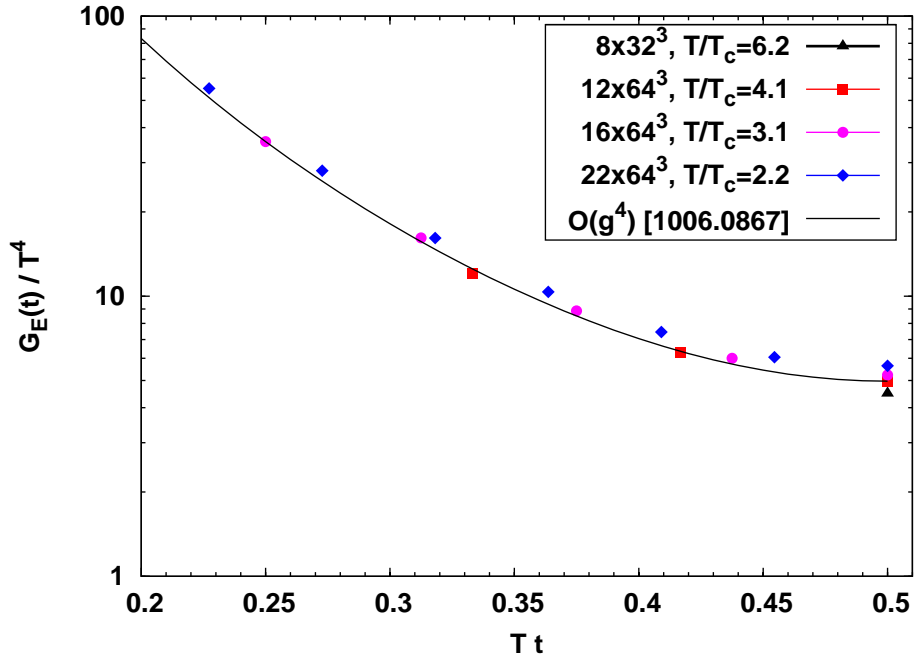}}
\caption{The Euclidean force-force correlator extracted from simulations 
with the Wilson action at $\beta=7.483$ and the HYP1 discretization of 
HQET. The black curve indicates the NLO result of Burnier et al~\cite{Burnier:2010rp}.
This result has been used to determine the normalization of the E-field
by matching the lattice data to it at $tT=\frac{1}{4}$ and $3.1T_c$. 
Statistical error bars are smaller than the data symbols.}
\label{fig:hyp}
\end{figure}

There is a lot of freedom in discretizing the HQET Lagrangian
(see~\cite{Sommer:2010ic} for an introduction).  As color parallel
transporters, we choose the lattice gauge links obtained after one
iteration of HYP-smearing~\cite{Hasenfratz:2001hp}. This action has
been used extensively at zero-temperature~\cite{Blossier:2010jk}, and
our choice of smearing parameters is known in the literature as the
{HYP1} action~\cite{DellaMorte:2005yc}. The original motivation for
choosing this action is that it reduces the size of the UV-divergent
self-energy, thus leading to a reduction in the statistical errors. At
the same time the cutoff effects were found to be
controllable. In~\cite{Guazzini:2007bu}, the chromo-magnetic field
$\boldsymbol{B}$ was also discretized with HYP-link variables,
yielding also a benefit in statistical error reduction.
Following this example, we use HYP-links for the chromo-electric field as well.

While the renormalization factor for the $\boldsymbol{B}$ field was
computed in~\cite{Guazzini:2007bu} (it is scale-dependent), the
corresponding factor for the chromo-electric field remains to be
computed. The same methods apply to the renormalization of this
factor, but in the mean time we use a preliminary way of normalizing
the chromo-electric operator. We expect that at short time separations
$t$, perturbation theory provides an accurate prediction for $G_E(t)$.
Therefore the absolute normalization of the chromo-electric field can
be obtained by requiring that the lattice correlator match the NLO
perturbative prediction at some reference time $t_{\rm ref}$.  The
difficulty in this procedure is that $t_{\rm ref}$ must still be large
enough in lattice units for the discretization errors to be under
control. In this work we made the compromise to choose $t_{\rm ref} =
1/4T$ at a temperature of about $3.1T_c$ on a $16\times64^3$ lattice.
In physical units, this represents a separation of about 0.05fm, and
four lattice spacings in lattice units.  At one value of the bare
coupling $g_0^2$, this normalization is then valid for other values of
$\Nt$. In this way, we have varied the temperature by changing $\Nt$
from 8 to 22 at fixed $6/g_0^2=7.483$. The chromo-electric field
correlators are displayed in \fig\ref{fig:hyp}.  Only those points
with $t/a\geq 4$ are displayed.  At shorter separations, the lattice
correlators exhibit non-monotonicity (and are even negative at
separation 0 and 1 lattice spacing), a not unexpected lattice
artefact.  It is clear that with the procedure adopted here, we cannot
disentangle discretization errors from the bare-coupling dependence of
the renormalization factor of $\boldsymbol{E}$. Our results should
accordingly be regarded as preliminary. Eventually the renormalization
factor should be computed along the lines of~\cite{Guazzini:2007bu},
where the $\boldsymbol{B}$ operator was treated instead.

\begin{figure}
\centerline{\includegraphics[width=8.7 cm,angle=-90]{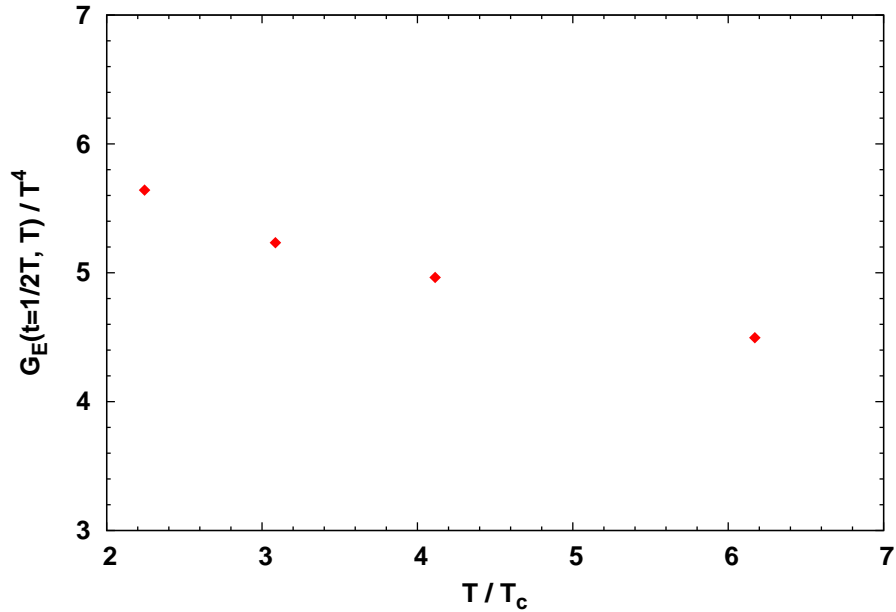}}
\caption{The Euclidean force-force correlator, normalized as in \fig\ref{fig:hyp}, at $t=1/2T$,
with $\beta=7.483$ and $\Nt=8,12,16,22$. Statistical error bars are smaller than the data symbols.}
\label{fig:midhyp}
\end{figure}

To repeat, the data at $3.1T_c$ has been calibrated to match the
O($g^4$) result of~\cite{Burnier:2010rp} at $t=1/4T$. In
\fig\ref{fig:hyp}, different temperatures are accessed by varying
$\Nt$. On the logarithmic scale of the plot, the temperature
dependence of the Euclidean correlator normalized by $T^4$ is
weak. Furthermore the NLO perturbative prediction provides a rather
good description of the $t$-dependence at $3.1T_c$.  The temperature
dependence of the mid-lattice data points (normalized as in
\fig\ref{fig:hyp}) is shown in much greater detail in
\fig\ref{fig:midhyp}.  The temperature variation between $2T_c$ and
$6T_c$ is on the order of $20\%$. The sign of the variation is the
same as predicted by the perturbative expression: the magnitude of the
force-force correlator in the gluon plasma increases as the
temperature decreases.

\begin{figure}
\centerline{\includegraphics[width=6.0 cm,angle=-90]{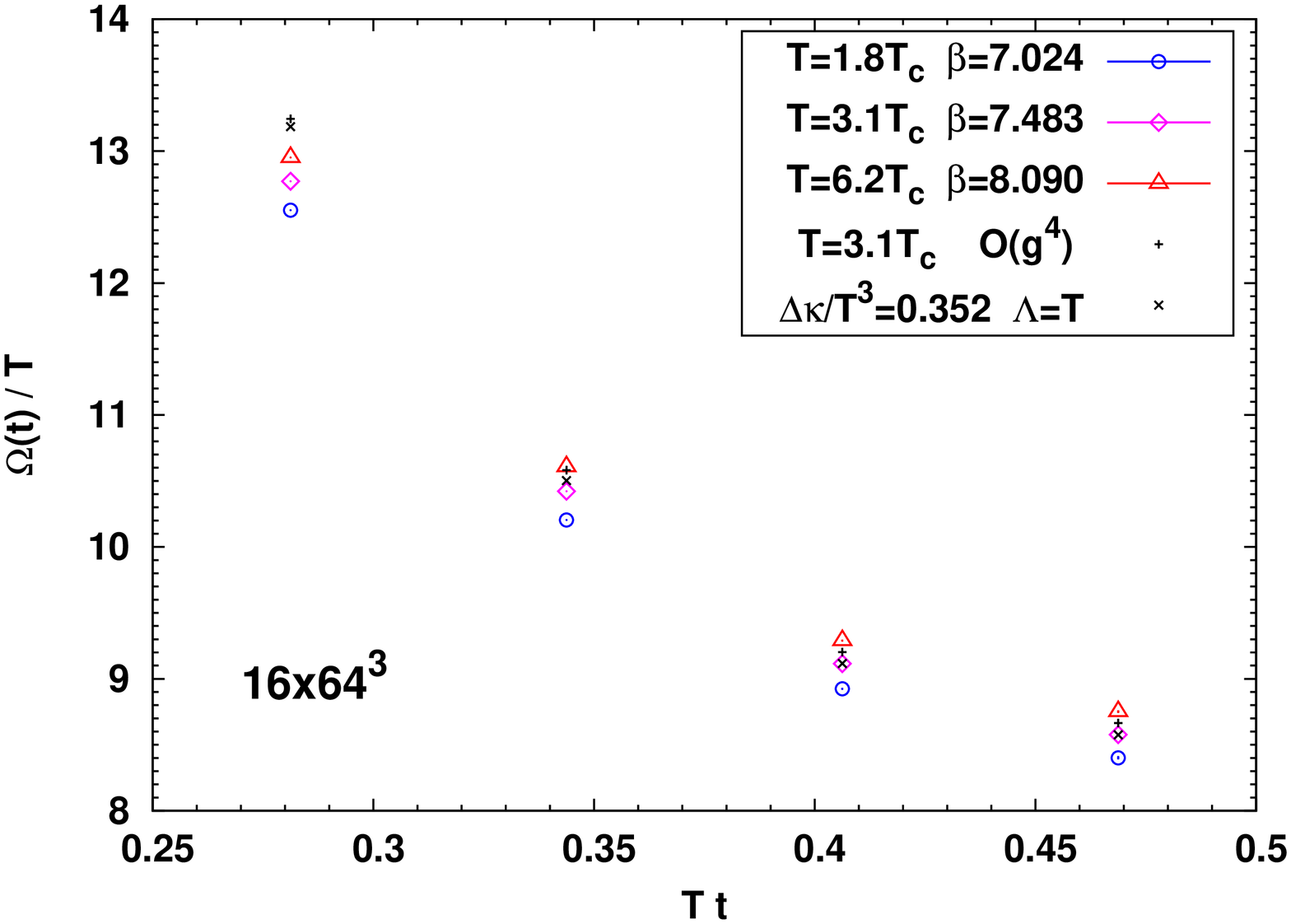}
\hspace{-0.7cm}
\includegraphics[width=6.0 cm,angle=-90]{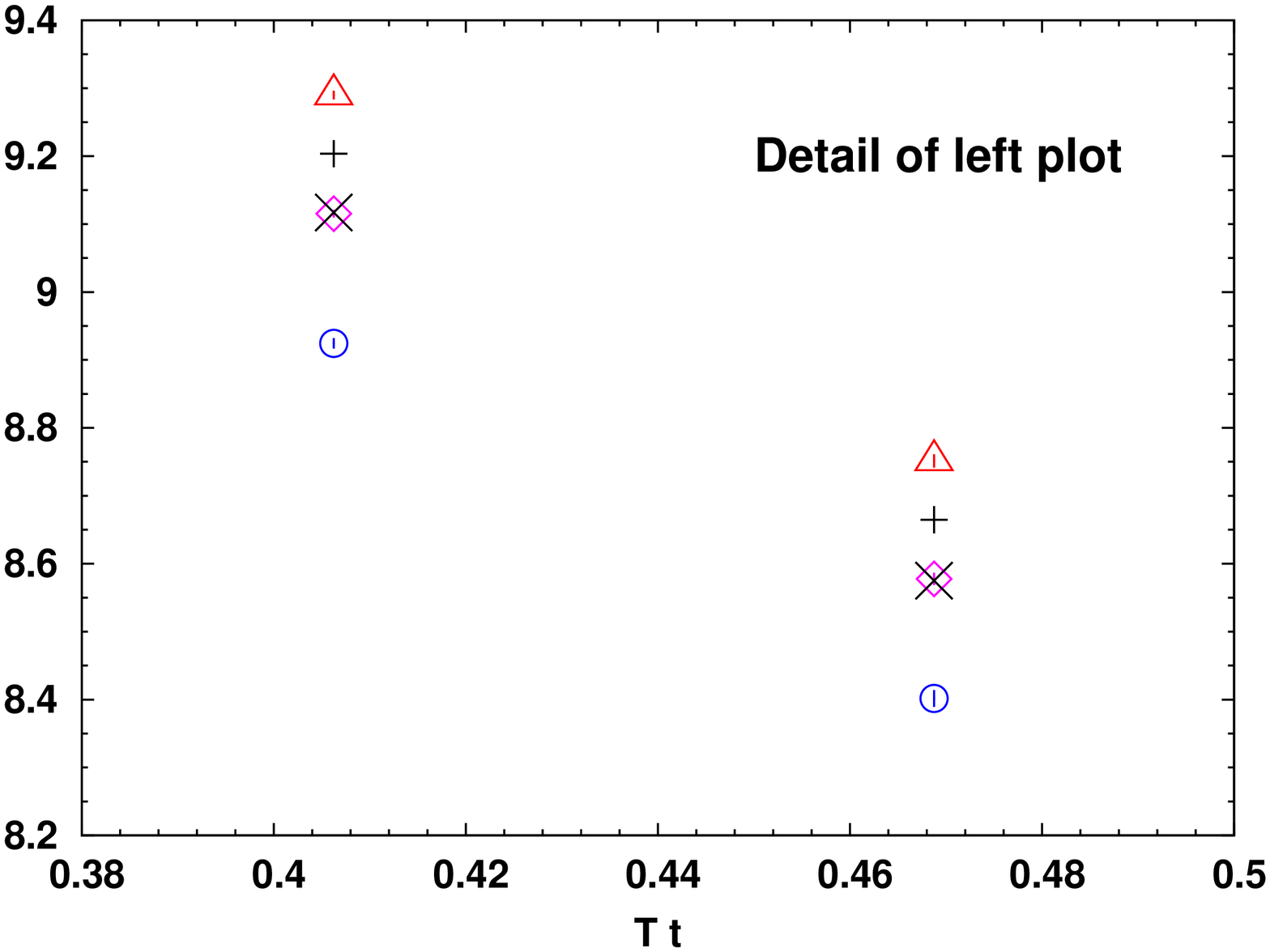}}
\caption{
The quantity displayed is $\Omega(t)$, defined in
\eq(\ref{eq:rat}). The `$+$' denote the $O(g^4)$ prediction at
$3T_c$~\cite{Burnier:2010rp}.  The `$\times$' denote the function
$\Omega(t)$ that results from adding the low-frequency contribution
$\Delta\rho(\omega)=\frac{1}{\pi}\Delta\kappa
\tanh(\frac{\omega}{2T})\theta(\Lambda-|\omega|)$ to the O($g^4$) spectral
function, with parameters $\Delta\kappa$ and $\Lambda$ given in the
caption.  Statistical error bars are smaller than the data symbols,
but discretization errors are probably non-negligible, particularly at
the smaller $t$ values.
}
\label{fig:hyp3Tc}
\end{figure}

Although we do not yet have non-perturbative control over the absolute
normalization of the correlators, the latter cancels out in 
the relative fall-off of the correlator. An observable that measures this is
the quantity $\Omega(t)\geq 0$ defined by 
\be
\frac{G_E(t-a/2)}{G_E(t+a/2)} =
\frac{\cosh\big[\Omega(t)({\beta}/{2}-(t-a/2))\big]}
     {\cosh\big[\Omega(t)({\beta}/{2}-(t+a/2))\big]}
\la{eq:rat}
\ee
We remark that $\Omega(t)$ has a continuum limit, in which $\Omega
\tanh\Omega(\beta/2-t)=-\frac{d}{dt}\log G_E(t)$. One can interpret it
as the location of a delta function in the spectral function which by
itself reproduces the local fall-off of the Euclidean correlator.  The
function $\Omega(t)$ is displayed in \fig\ref{fig:hyp3Tc} for three
different temperatures, where the temperature is varied this time by
changing the bare coupling $g_0^2$ at fixed $\Nt=16$.  We note that
because the statistical samples of the numerator and the denominator
in \eq(\ref{eq:rat}) are highly correlated, the numerical results for
these ratios have uncertainties at the few-permille level. Figure
\ref{fig:hyp3Tc} also displays the perturbative prediction at $3.1T_c$
corresponding to the curve appearing in \fig\ref{fig:hyp}; it is based
directly on \eq(\ref{eq:rat}) rather than on the continuum version of
this equation, to allow for a more direct comparison with the lattice
data. The lattice $\Omega(t)$ differs from the perturbative one only
by a few percent, namely the latter falls off slightly more steeply.
Since the difference is so small, it could partly be due to
discretization effects. To reduce their influence, we concentrate on
the largest values of $t$.  One may ask nonetheless, how large a
difference in the transport coefficient $\kappa$ could this
discrepancy possibly correspond to. An estimate is obtained by adding
a low-frequency correction to the perturbative spectral function,
\be
\Delta\rho(\omega)=                                                                     
\frac{1}{\pi}\cdot\Delta\kappa\, 
\tanh({\txts\frac{\omega}{2T}})\;\theta(\Lambda-|\omega|).
\la{eq:Drho}
\ee
Other functional forms than (\ref{eq:Drho}), such as a Breit-Wigner
curve, would perhaps be more realistic, but would not change our
conclusions in any significant way.  Adding such a term to the
spectral function has the effect of making the Euclidean correlator
flatter, and indeed, by adjusting $\Delta\kappa$, one can obtain good
agreement for the largest two $t$ values between the perturbative
prediction modified by \eq(\ref{eq:Drho}) and the lattice data.  At
$T=3.1T_c$ we find that, for $\Lambda=T$ and $\Delta\kappa/T^3= 0.352(38)$,
agreement is obtained with the lattice data at the two largest $t$ values. 
This represents a substantial enhancement of $\kappa$ over
the leading-order perturbative value mentioned in the introduction.
An equally good agreement is obtained if one chooses $\Lambda=2T$, 
which leads to $\Delta\kappa/T^3\approx0.204(22)$.

\section{Conclusion}

To summarize, we have found that the Euclidean force-force correlator
evaluated at $t={\beta/2}$ admits a $\approx20\%$ increase as the
temperature is lowered from 6 to $2T_c$.  At $3.1T_c$ its
$t$-dependence is described at the few-percent level by the recent NLO
perturbative result~\cite{Burnier:2010rp}. The somewhat flatter
behavior seen in the lattice data can be explained by an enhancement
of the spectral function at low frequencies, and adopting this explanation
leads to a substantial increase of $\kappa$ over the leading-order result obtained
in~\cite{Moore:2004tg}.  While it is too early to draw
phenomenological conclusions, at present the increase appears to be not quite
sufficient to explain the experimentally observed elliptic flow of
heavy quarks, as discussed in the introduction. It will be interesting
to see the results brought by the current lead-lead collisions at the
LHC. The ALICE experiment has very recently
reported~\cite{Aamodt:2010pa} an integrated elliptic flow (i.e.~of
light constituents) about $30\%$ larger than at RHIC.

From a technical point of view, it is encouraging that a precision at the few
per-mille level can be achieved on the force-force correlator on a
$16\times64^3$ lattice.  Given this statistical precision, it is
essential, as a next step, to control the discretization errors at a
comparable level. One could then study more systematically and
quantitatively the implications of the lattice data for the spectral
function. A mandatory step in controlling the discretization errors is
to compute the renormalization factor of the $\boldsymbol{E}$ field
non-perturbatively. The motivation to carry out these calculations is now 
quite strong.

\acknowledgments{
I thank Mikko Laine for interesting discussions and for providing the 
O($g^4$) prediction in \fig\ref{fig:hyp}, and Rainer Sommer 
for helpful discussions about HQET.
Lattice computations for this work were  carried out on facilities of
the USQCD Collaboration, which are funded by the Office of Science of
the U.S. Department of Energy.
}

\appendix
\section{Thermal correlators and linear response}
We start by recalling some of the definitions, which allows us 
to fix our notation.
At finite temperature $T\equiv1/\beta$, correlation functions are defined as
\be
G_>^{AB}(t)\equiv \tr\{ \hat\rho A(t) B(0) \}\,,
\la{eq:G>}
\ee
with $\hat\rho \equiv \frac{1}{Z} e^{-\beta H}$ the equilibrium density matrix.
Expectation values of commutators,
\be
G^{AB}(t) = i \tr\{\rho [A(t),B(0)]\} = i\left(G_>^{AB}(t) - G_>^{BA}(-t)\right),
\la{eq:G}
\ee
play a particularly important role in finite-temperature physics.
The integral transform over the positive half-axis
\be
G_R^{AB}(\omega) = \int_0^\infty \ud t\, e^{i\omega t} G^{AB}(t)
\ee
is analytic in the half complex plane $\im(\omega) > 0$.
We will refer to it as the retarded correlator.
The spectral function, which is really a distribution, is defined as 
\be
\rho^{AB}(\omega)= \frac{1}{2\pi i}
\int_{-\infty}^{+\infty} \ud t\; e^{i\omega t}\; G^{AB}(t).
\la{eq:rhodef}
\ee
For $B=A^\dagger$, the spectral function is identically related 
to the imaginary part of the retarded correlator,
\be
\rho^{AA^\dagger}(\omega) = \frac{1}{\pi} \im G^{AA^\dagger}_R(\omega)\in \mathbb{R}\,.
\la{eq:rho-ImGR}
\ee
The Euclidean correlator is defined as
\be
G_E^{AB}(t) = G_{>}^{AB}(-it).
\la{eq:GE}
\ee
It obeys the Kubo--Martin-Schwinger relation 
\be
G_E^{BA}(\beta-t) = G_E^{AB}(t)\,,
\ee
and therefore admits a representation as a Fourier series,
\be
G_E(t) = T\sum_{\ell\in{\mathbb{ Z}}} G_E^{(\ell)}\, e^{-i\omega_\ell t}\,,
\qquad\quad
G_E^{(\ell)} =  \int_0^\beta  \ud t\, e^{i\omega_\ell t} G_E(t)\,,
\la{eq:GEell}
\ee
where $\omega_\ell  = 2\pi T\,\ell$ and we have dropped 
the label specifying the operators $A, B$.

In frequency space, the Euclidean and retarded correlators are related by 
\be
G_R(i\omega_\ell) = G_E^{(\ell)} \,.
\la{eq:l}
\ee
The $\ell=0$ case has to be treated somewhat more carefully, see 
\cite{Meyer:2010ii} for details.
In coordinate space, the relation between the Euclidean correlator and
the real-time correlator is
\ba
G^{AB}_E(t) + G^{BA}_E(t) = T\int_0^\infty \!\!\!\ud t'\, (G^{AB}(t') +G^{BA}(t'))\, 
\frac{\sinh(2\pi T t')}{\cosh(2\pi Tt')-\cos(2\pi Tt)},
\la{eq:GEG}
\\
i\big(G^{AB}_E(t) - G^{BA}_E(t)\big) = T\int_0^\infty \!\!\!\ud t'\, (G^{AB}(t') - G^{BA}(t'))\, 
\frac{\sin(2\pi T t)}{\cosh(2\pi Tt')-\cos(2\pi Tt)}.
\ea
Finally, the most commonly used relation between Euclidean and real-time correlators is 
the mixed coordinate-frequency space relation
\ba
G^{AB}_E(t) + G_E^{AB}(\beta-t)
&=& \!\int_{-\infty}^\infty \!\!\!\!\ud\omega\, \rho^{AB}(\omega)\,
\frac{\cosh\omega(\frac{\beta}{2}-t)}{\sinh\beta\omega/2},
\la{eq:ClatRho}
\\
G^{AB}_E(t) - G_E^{AB}(\beta-t)
&=&  \!\int_{-\infty}^\infty\!\!\!\! \ud\omega\, \rho^{AB}(\omega)\,
\frac{\sinh\omega(\frac{\beta}{2}-t)}{\sinh\beta\omega/2}.
\ea

The retarded correlator $G_R^{AB}$ is important because it is related to the 
response of operator $A$ to a perturbation of the system by operator $B$.
A time-dependent perturbation of the Hamiltonian by an operator $B$,
\be
H_f(t) = H - f(t) B(t),
\la{eq:Hpert}
\ee
leads to a `response' of physical quantities, i.e.~a change 
in their expectation values with respect to the unperturbed ensemble.
The evolution equation of an operator $A$ is given by
\be
i\frac{\partial}{\partial t}A(t) = -[H_f(t),A(t)].
\ee
One then finds that to linear order in $f$, the expectation value of $A$ 
in the perturbed system is 
\be
\delta\<A(t)\> \equiv  \<A(t)\>_f - \<A(0)\>
\la{eq:linresp1}
= \!\int_{-\infty}^t \!\ud t' G^{AB}(t-t') f(t')+ {\rm O}(f^2).
\ee
Equation (\ref{eq:linresp1}) is the master formula of linear response
theory. It shows that the retarded correlator $G_R^{AB}$ determines
the response of an observable $A$ to a time-dependent external field
that couples to $B$.
A source term of the form
\be
f(t) = e^{\epsilon t} \theta(-t) f_0
\la{eq:adiabpert}
\ee
is often adopted to study how the system relaxes back to equilibrium
after having been perturbed adiabatically.
The static susceptibility is defined as the expectation value of $A$ at $t=0$,
\be
\delta\<A(t=0)\>_f = \chi^{AB}_s \, f_0.
\la{eq:chi_stat1}
\ee
From (\ref{eq:linresp1}), it follows that 
\be
\chi^{AB}_s = \int_0^\infty \ud t\, e^{-\epsilon t} \,G^{AB}(t)\,
= G^{AB}_R(i\epsilon).
\la{eq:chi_stat2}
\ee
Integrating both sides of (\ref{eq:linresp1}), $\int_0^\infty
\ud\omega\,e^{i\omega t}(.)$, one obtains for the adiabatic
perturbation (\ref{eq:adiabpert}) 
\be G_R^{AB}(\omega)f_0 = \<\delta
A(0)\>_f + i\omega \int_0^\infty \ud t \, e^{i\omega t} \<\delta
A(t)\>_f.  
\la{eq:retresp} 
\ee 
This formula shows that the relaxation of observable $A$ back to its
equilibrium value and the retarded correlator $G_R^{AB}(\omega)$ are
in one-to-one correspondence. Since the late-time relaxation is
expected to be described by hydrodynamic evolution, this equation can
be exploited to obtain a prediction of the small-$\omega$ functional
form of $G_R$.

\bibliography{/home/meyerh/CTPHOPPER/ctphopper-home/BIBLIO/viscobib.bib}

\end{document}